\documentclass[conference]{IEEEtran}
\IEEEoverridecommandlockouts
\usepackage{amsmath,amssymb,amsfonts}
\usepackage{booktabs} 
\usepackage{graphicx}
\usepackage{balance}
\usepackage{pifont} 
\usepackage{longtable}
\usepackage{enumitem} 
\usepackage[algoruled, linesnumbered, vlined]{algorithm2e}
\usepackage{setspace}
\usepackage{multirow}
\usepackage{url}
\usepackage[T1]{fontenc}
\usepackage{hyperref}
\usepackage{romannum}

\usepackage{sparklines}
\usepackage[export]{adjustbox}
\usepackage{subfig}
\usepackage{xspace}
\usepackage{xcolor}
\usepackage{soul}
\usepackage{marginnote}
\newcommand{\ie}{i.e.,\ }
\newcommand{\eg}{e.g.,\ }

\definecolor{coolblack}{rgb}{0.0, 0.18, 0.39}

\newcommand{\tinyskip}{\vspace{3pt}}
\newcommand{\mypar}[1]{\tinyskip\noindent\textbf{#1.}\xspace}

\newcommand{\F}{\mbox{Fig.\hspace{0.25em}}}

\newenvironment{myitemize}{%
\begin{itemize}[leftmargin=1em, itemsep=.1em, parsep=.1em, topsep=.1em,
    partopsep=.1em]}
{\end{itemize}}

\newenvironment{myenumerate}{%
\begin{enumerate}[leftmargin=1em, itemsep=.1em, parsep=.1em, topsep=.1em,
    partopsep=.1em]}
{\end{enumerate}}

\newenvironment{structure*}{\color{blue}\begin{myenumerate}}{\end{myenumerate}}

\sethlcolor{yellow}

\DeclareMathOperator*{\argmax}{argmax}

\def\BibTeX{{\rm B\kern-.05em{\sc i\kern-.025em b}\kern-.08em
    T\kern-.1667em\lower.7ex\hbox{E}\kern-.125emX}}
\begin{document}

\title{FabricQA-Extractor: A Question Answering System to Extract Information from Documents using Natural Language Questions}


\author{\IEEEauthorblockN{Qiming Wang}
\IEEEauthorblockA{\textit{Department of Computer Science} \\
\textit{University of Chicago}\\
Chicago, USA \\
qmwang@uchicago.edu}
\and
\IEEEauthorblockN{Raul Castro Fernandez}
\IEEEauthorblockA{\textit{Department of Computer Science} \\
\textit{University of Chicago}\\
Chicago, USA \\
raulcf@uchicago.edu}
}

\maketitle

\begin{abstract}

Reading comprehension models answer questions posed in natural language when
provided with a short passage of text.  They present an opportunity to address a
long-standing challenge in data management: the extraction of structured data
from unstructured text.  Consequently, several approaches are using these models
to perform information extraction. However, these modern approaches leave an
opportunity behind because they do not exploit the \emph{relational} structure of the
target extraction table.

In this paper, we introduce a new model, Relation Coherence, that exploits
knowledge of the relational structure to improve the extraction quality. We
incorporate the Relation Coherence model as part of FabricQA-Extractor, an
end-to-end system we built from scratch to conduct large scale extraction tasks
over millions of documents. We demonstrate on two datasets with millions of
passages that Relation Coherence boosts extraction performance and evaluate
FabricQA-Extractor on large scale datasets. 


%

\end{abstract}


\section{Introduction}

A long-standing challenge in data management is the extraction of knowledge from
unstructured sources, such as text. Data management techniques have perfected
techniques to query relations and other structured data, but they do not work as
well on text and other unstructured data. Because this leaves a vast amount of
knowledge untapped, multiple communities of academics and practitioners use
\emph{information extraction} techniques to structure unstructured data, thus,
making it queryable with data management systems. In databases, pioneering work
in information extraction such as the Knowitall~\cite{knowitall} system has been
continued more recently with efforts such as Deepdive~\cite{deepdive}.  Besides
databases, other academic communities---notably NLP and IR---continue to
improve the \emph{information extraction pipeline}, referring to the problem in
various ways that include automatic knowledge base completion~\cite{kbp-approch,
tac-kbp-heng, tac-kbp-2}, slot-filling tasks~\cite{pasd-sf, stanford-2014-sf,
unist-tac-2017, stanford-tac-2016, stanford-tac-2017}, and others.

Creating information extraction pipelines is a labor-intensive task, even when
using state of the art technology. First, it requires training data for each
relation of interest. For example, to extract instances of \emph{born\_in} in
new text, systems such as Knowitall\cite{knowitall}, Deepdive\cite{deepdive} need
training data that presents example extractions. Second, a machine learning
model is built and tuned for each relation. These two steps must be repeated for
each relation of interest. Although techniques such as distant
supervision~\cite{distantsupervision} and advances in automatically labeling
data~\cite{snorkel} have improved dramatically over the last few years, the
process remains time-consuming and available only to teams with access to high
expertise. For this reason, many modern information extraction approaches rely
on other sources of external knowledge, such as \cite{tac-kbp-heng, kbpearl}. We focus on scenarios without that external knowledge.

\begin{figure}[t] 
\centering
\includegraphics[width=\columnwidth]{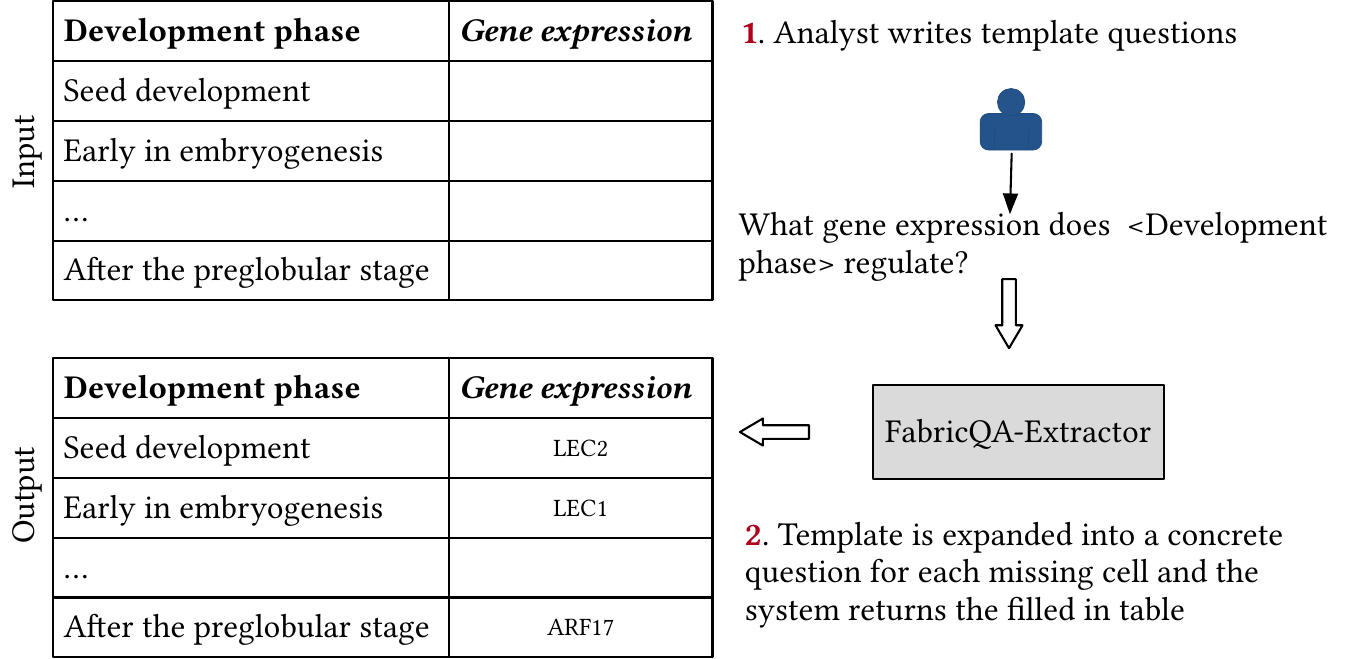}
\caption{FabricQA-Extractor user interface.} 
\label{fig:diagram} 
\end{figure}

\mypar{An opportunity} Recent trends in natural language processing open new
opportunities to extract information from text without relation-specific machine
learning models. For example, reading comprehension models~\cite{albert, bidaf}
(RC models) provide the answer to a natural language question when a short
passage of text is provided as input. And these models have been incorporated in
open domain question answering systems that, given a large collection of
documents,
identify first the right passage, and then feed the passage to the RC model to
obtain the answer~\cite{drqa}. But these approaches are generally designed to
answer natural questions, and not to extract structured information from text.
When extracting information from text we have access to two additional sources
of information that question answering systems do not exploit: i) we know the
target schema (\eg see \F\ref{fig:diagram}); and ii) we
know that the relationship between elements in rows is the same for each row in
the target relation.

In this paper, we make two contributions. The main contribution is a new
\textbf{Relation Coherence} model that exploits knowledge about the relational
model to augment the information available to open question answering systems,
resulting in a boost in the extraction quality. The second contribution is a
system, \textbf{FabricQA-Extractor}. FabricQA-Extractor is built as a sequence
of increasingly more sophisticated components that filter out millions of
input documents into 1 answer per missing cell in the input table. This
\emph{funnel design} fills in each cell in sub-second latencies, making
the whole system practical for large scale information extraction tasks. In
summary, the new Relation Coherence model together with FabricQA-Extractor reduce
the effort of extracting data from large collections of documents.




\mypar{Query Lifecycle and Example} \emph{Consider an analyst who wants to
identify what gene expression regulates a list of development phases.
Conceptually, the knowledge the analyst has is represented in table
~\ref{table:landscape} with the filled-in cells (see \F\ref{fig:diagram}). Empty
cells represent the information the analyst wants to identify. Traditionally,
the analyst would need to search over many documents to find the answers.} 

With today's information extraction systems, the analyst needs to train and tune
pipelines for each relation of interest, \eg \emph{Development phase }and
\emph{Gene expression}. Instead, with FabricQA-Extractor, the analyst
only has to write one question template, such as the one shown at the top of
\F\ref{fig:diagram} and the system populates the table with answers from the
documents.

To achieve that, FabricQA-Extractor ingests and chunks a collection of
documents that are made searchable to a pipeline of stages. During the first
stage, a Passage Ranker selects a subset of promising passages, which
are sent to an Answer Ranker. This, in turn, extracts the candidate
answers, which are forwarded to the Relation Coherence model for the
final selection. The Relation Coherence knows that if there is a
relationship between \emph{Development Phase} and \emph{Gene Expression}, there
is also an inverse relationship between \emph{Gene Expression} and
\emph{Development Phase},  and that these relationships are the same across
rows of the relation. It uses this knowledge to augment the information
available to the system. At the end of the process, each missing cell in the
input table contains an answer. Analysts can inspect the cells to get access to
a top-k answers, each accompanied of useful provenance, such as the passage
where it was extracted from, and the document that contains that passage. This
provenance information helps analysts interpret the results of the system.



\mypar{Evaluation} We demonstrate on two datasets with tens of millions of passages, QA-ZRE and
BioNLP, that the Relation Coherence model
improves the quality extraction of large-scale information extraction tasks. We
use standard benchmarks for question answering over passages, but execute them
on a much more challenging scenario: the input to our system is not a passage of
text, but a collection of documents with tens of millions of such passages. We
show that the Relation Coherence model performs well even when changing
the domain of the documents to biomedical text for which the model was not
trained. We evaluate the end-to-end design of FabricQA-Extractor by
measuring its extraction throughout. We demonstrate the importance of carefully
designing the non-ML-based components.  For example, a good document indexing
strategy boosts the end-to-end quality of the extraction as much as improvements
on the ML models themselves. 

%

The rest of the paper is structured as follows. We present the background in
Section~\ref{sec:background}. The Relation Coherence model is presented in
Section~\ref{sec:relver} and the FabricQA-Extractor system in
Section~\ref{sec:fullsystem}. The evaluation results
(Section~\ref{sec:evaluation}) are followed by related work
(Section~\ref{sec:relatedwork}) and conclusions in
Section~\ref{sec:conclusions}.


\section{Preliminaries}
\label{sec:background}

In this section, we describe the problem setup in Section~\ref{subsec:setting},
present a solution landscape that includes techniques in information extraction,
reading comprehension, and question answering systems in
Section~\ref{subsec:landscape}. We finish presenting a summary and problem
statement in Section~\ref{subsec:problemstatement}.

\subsection{Problem Setup}
\label{subsec:setting}

We consider partially filled relations with $R$ attributes, one of which is the
\emph{key attribute}, $R_k$. An example is shown in \F\ref{fig:diagram}.  Each
cell in the table can be represented with the triple $(subject,
relationship, object) (S, R, O)$. A missing cell is represented as a
triple that lacks either the subject (when the value falls under $R_k$)
or the object. Given one such partially filled relation, the goal is to
fill in the missing cells using information from text documents.  

%

One way of filling the missing cells in the partially filled table is to
decompose it into multiple slot-filling tasks, each in charge of filling an
empty cell, \ie triple. Then, we can leverage the vast body of work in
slot-filling tasks~\cite{pasd-sf, stanford-2014-sf, unist-tac-2017,
stanford-tac-2016, stanford-tac-2017, openie-tac-2017} automatic knowledge base
completion (AKBC)~\cite{kbc-qa}, and open information
extraction~\cite{openie-6}. We analyze techniques to solve this problem in the next section.

\mypar{Imputation Methods} With a partially filled table, it is natural to
consider imputation methods for dealing with missing data, as in the data
quality literature. Traditional filling missing values problems such as
\cite{netflix_prize, lessons_netflix_prize, bellkor_solution,
Hamilton_graph_learning, Kipf_gcnn, lu_lpcn, Teru_graph_rel} usually rely on
statistical techniques and knowledge from neighboring cells that are used to
compute the missing cells. Unlike these, we target scenarios where
the answer cannot be computed out of the existing information in the table.
Instead, we want to populate the table by finding the values for the missing
cells among a collection of documents. 

\subsection{Solution Landscape}
\label{subsec:landscape}

We describe three general approaches, \emph{information extraction},
\emph{reading comprehension}, \emph{open domain question answering}, in
Table~\ref{table:landscape}. 

\begin{table*}[ht]
\center
\scalebox{0.9}{
\begin{tabular}{|c|c|c|c|c|c|c|}
\hline
\multirow{2}{*}{\textbf{Category}}                                                       & \multirow{2}{*}{\textbf{Approach}} & \multirow{2}{*}{\textbf{\begin{tabular}[c]{@{}c@{}}Designed for \\ Large Scale\end{tabular}}} & \multicolumn{2}{c|}{\textbf{User Effort}}                                                                                                                                          & \multirow{2}{*}{\textbf{\begin{tabular}[c]{@{}c@{}}Model vs \\ Full System\end{tabular}}} & \multirow{2}{*}{\textbf{\begin{tabular}[c]{@{}c@{}}Exploit \\ Relation Model\end{tabular}}} \\ \cline{4-5}
                                                                                         &                                    &                                                                                               & \textbf{\begin{tabular}[c]{@{}c@{}}Requires \\ Annotating \\ Corpus\end{tabular}} & \textbf{\begin{tabular}[c]{@{}c@{}}Requires Training \\ Dataset per \\ Relationship\end{tabular}} &                                                                                           &                                                                                             \\ \hline
\multirow{6}{*}{\begin{tabular}[c]{@{}c@{}}Relation extraction \\ using QA\end{tabular}} & Levy et. al, 2017\cite{zero-shot-rel}                  & N                                                                                             & N                                                                                 & N                                                                                              & Model                                                                                     & N                                                                                           \\ \cline{2-7} 
                                                                                         & Li et. al, 2019\cite{li2019entity}                    & N                                                                                             & N                                                                                 & N                                                                                              & Model                                                                                     & N                                                                                           \\ \cline{2-7} 
                                                                                         & Zhao et. al, 2020\cite{zhao2020asking}                 & N                                                                                             & N                                                                                 & Y                                                                                              & Model                                                                                     & N                                                                                           \\ \cline{2-7} 
                                                                                         & West et. al, 2014\cite{kbc-qa}                  & Y                                                                                             & N                                                                                 & Y                                                                                              & System                                                                                    & N                                                                                           \\ \cline{2-7} 
                                                                                         & Qiu et. al, 2018\cite{qiu2018qa4ie}                   & N                                                                                             & Y                                                                                 & N                                                                                              & System                                                                                    & N                                                                                           \\ \cline{2-7} 
                                                                                         & Das et. al, 2018\cite{das2018building}                   & N                                                                                             & N                                                                                 & Y                                                                                              & Model                                                                                     & N                                                                                           \\ \hline
\multirow{2}{*}{\begin{tabular}[c]{@{}c@{}}Non-neural \\ slot filling\end{tabular}}      & Angeli et. al, 2014\cite{stanford-2014-sf}                 & N                                                                                             & N                                                                                 & Y                                                                                              & System                                                                                    & N                                                                                           \\ \cline{2-7} 
                                                                                         & Roth et. al, 2014\cite{roth2014effective}                  & Y                                                                                             & N                                                                                 & Y                                                                                              & System                                                                                    & N                                                                                           \\ \hline
\begin{tabular}[c]{@{}c@{}}Neural \\ slot filling\end{tabular}                           & Zhang et. al, 2017\cite{pasd-sf}                 & N                                                                                             & N                                                                                 & Y                                                                                              & Model                                                                                     & N                                                                                           \\ \hline
\multirow{4}{*}{OpenQA}                                                                  & DRQA, 2017)\cite{drqa}                        & Y                                                                                             & N                                                                                 & N                                                                                              & System                                                                                    & N                                                                                           \\ \cline{2-7} 
                                                                                         & PathRetriever, 2020\cite{asai2020learning}               & Y                                                                                             & Y                                                                                 & N                                                                                              & Model                                                                                     & N                                                                                  \\ \cline{2-7} 
                                                                                         & SPARTA, 2021\cite{zhao2020sparta}                      & Y                                                                                             & Y                                                                                 & N                                                                                              & Model                                                                                     & N                                                                                           \\ \cline{2-7} 
                                                                                         & MDR, 2021\cite{xiong2020answering}                         & Y                                                                                             & N                                                                                 & N                                                                                              & Model                                                                                     & N                                                                                           \\ \hline
                                                                                         & FabricQA-Extractor                 & Y                                                                                             & N                                                                                 & N                                                                                              & System                                                                                    & \textbf{Y}                                                                                  \\ \hline
\end{tabular}
}
\caption{Landscape of approaches related to information extraction and question
answering.}
\label{table:landscape}
\end{table*}

\subsubsection{Information Extraction}

Information extraction (IE) techniques can be broadly classified into 3
categories:

\noindent$\bullet$ \emph{Open IE}. The goal is to extract $(S, R, O)$ triples
from a collection of documents.

\noindent$\bullet$ \emph{IE over relations of interest}. Given a relation of
interest, $R_i$, the goal
is to extract all $(S, R_i, O)$ triples from the collection of documents.

\noindent$\bullet$ \emph{Slot filling (SF)}. Given $S_i$ and $R_i$, the goal
is to extract all $(S_i, R_i, O)$ triples from the collection of documents.

Given a partially filled relation, we are concerned with the third category.
Many SF techniques build ML
models to predict $O$ from training datasets that consist of 
sentences where $S_i$, $R_i$ and $O$ co-occur.
After the model is trained on that corpus, it can be used to predict $O$ given
the $S_i$ and $R_i$ for which it was trained. Many IE techniques \cite{slp, nlp,
deepdive, roth2014effective, stanford-2014-sf} rely on handcrafted rules and
machine learning classifiers to identify the correct object from the text.
Neural relation extraction models \cite{pasd-sf, kumar2017survey} learn
linguistic features automatically that reduce the need for handcrafted rules,
but they still need a large amount of training data per relationship.

%

\mypar{Challenge} A partially filled table may have multiple different
attributes, $R$. The major impediment of this class of information extraction
techniques is that \emph{they require assembling a training dataset and training
a model} for each $R_i$ of interest. For example, an analyst aiming to extract
multiple tables will have to prepare a machine learning task for each attribute
in each table. As a consequence, these approaches require high user effort, as
indicated in Table~\ref{table:landscape}. 

\subsubsection{Reading Comprehension}
\label{subsubsec:rc}

Recent advances in \emph{machine reading comprehension} (RC) \cite{squad-1.0,
squad-2.0} from the natural language processing (NLP) community shed some new
light on how to fill in missing values for partially filled relations. State of
the art machine reader models such as Albert~\cite{albert} and
BiDAF~\cite{bidaf} take a short, 3- to 4-sentence passage of text and a question
in natural language and find the span of text in the input snippet that answers
the question. The last few years have seen an arms race
in improving the accuracy of these models, fueled in part by the popularity of
benchmarks such as SQuAD~\cite{squad-1.0}.

Consider how these models could be used to address the challenge above. A human
operator could provide the RC model with a natural question whose answer
corresponds to the missing cell and a passage that contains the answer. For
example, to fill in the cell that corresponds to the \emph{Gene expression}
regulated by \emph{Seed development}, the operator could ask: \emph{``What gene
expression does seed development regulate?''} and provide a paragraph with the
information.
The RC model would return the answer (the object in the triple), which
would fill in the blank. This promising approach is used by approaches such as
Zero-Shot relation extraction~\cite{zero-shot-rel} to address slot filling
tasks. 



\mypar{Challenge} The major challenge of using RC models is that they require
\emph{the right passage} as input, but we do not know a priori which passage
among the possibly millions of documents is relevant. This is represented in the
column ``Designed for Large Scale'' in Table~\ref{table:landscape}, where we
separate solutions that are designed to answer questions over passages from
those designed to answer questions over collections of documents.

Modern RC models can refuse to answer a question if they do not find a suitable
candidate in the provided passage, so in principle, we could split the documents
into multiple passages and fill in each one to the model, until we obtain an
answer. This solution would not work for two reasons. First, RC models are not
perfect, they will make mistakes, and the number of mistakes will be large when
the number of passages is large.  Second, and worse, running inference on each
passage of the input documents is not a scalable solution.
Performance-optimized RC models, such as Microsoft's version of
BERT-SQUAD~\cite{microsoft_bert_squad} have an average inference time of 1.7ms
per question. Even if we only consider 1 million passages, this results in 30
minutes per question, \ie 30 minutes to find answers for one single missing
cell. The scenarios we consider have many more times that number of passages
$\sim$30M, we want to fill in tables with hundreds of rows, and state of the art RC
models such as \cite{bert, albert} have slower inference time than Microsoft's
optimized version. 

\subsubsection{Open Domain Question Answering}

Because of the challenges of using RC models for large scale extractions, state
of the art open question answering systems (OpenQA) \cite{drqa, wang2018r, yang2019end, lee-etal-2019-latent, zhao2020sparta, asai2020learning, xiong2020answering, multi_passage_bert} use
an information retrieval (IR) system to narrow down the search space of
paragraphs and then apply the expensive RC models. Unfortunately, simply
combining an RC reader to a IR system misses the opportunity of leveraging the
downstream information provided by the table we are filling. 
A table involves multiple entities connected with different
relationships. Existing OpenQA system model the answerability of a question
given a passage but do not explicitly verify that the relationship is implied in
the passage. In addition, there are multiple tuples that share the same
relationship in the table, which can also be leveraged to improve the extraction
quality.

\subsection{Summary and Contribution}
\label{subsec:problemstatement}

\mypar{Summary} Traditional IE approaches are limited by 
the
need to train a classifier for each target relationship with the
consequent large user effort. RC models do not need to train for each new
relationship, but they are limited by the need of providing the correct passage.
Modern open question answering systems leverage RC models to answer questions
over collections of documents but do not exploit information from the relational
model. 

\mypar{Contributions} Relation Coherence uses the available target schema to
augment the amount of information available to open domain question answering
systems. We explain the model in detail in Section~\ref{sec:relver}.
FabricQA-Extractor is a full end-to-end system which we have built to operate
over large scale collections of documents while delivering subsecond latency
answers with high extraction quality. This is in contrast to many approaches in
the literature that only provide a model and do not solve the full end-to-end
engineering challenge. We reflect this characteristic using ``System'' in column
``Model vs Full System'' of Table~\ref{table:landscape}. We present
FabricQA-Extractor in Section~\ref{sec:fullsystem}.


\section{Relation Coherence Model}
\label{sec:relver}

In this section, we introduce the Relation Coherence model, which
exploits knowledge about the relational model in order to improve extraction
quality.

\mypar{High-level intuition} Given a subject $S$, relation $R$, and a question $Q$, we
say a passage is correct if it contains $S$, $R$, and crucially, $O$, which corresponds
to the answer to $Q$. Then, when we ask $Q$, we look for passages with $S$ and $R$ that may
contain $O$. The insight we introduce in this paper is that given $R$, $O$, and a
reverse question $\overleftarrow{Q}$, a correct passage should contain $S$ as well.
This is a direct result of having access to the target relation, and hence,
knowing $S$, and $R$. The Relation Coherence model uses this additional information
to score candidate passages in a way that boosts the performance of traditional
OpenQA systems. In other words, while today's OpenQA systems only use the score
given by the reading comprehension model to determine whether a passage likely
contains an answer to a question, the Relation Coherence model augments that
score with the opposite task: whether a passage contains the $S$ for a $R$, $O$. The
resulting coherence score results in a ranking that boosts the extraction
quality.

We call Forward Searching to the process of identifying $O$ given $S$ and $R$:
this is the task that OpenQA systems perform. We call Backward Searching
(Section~\ref{subsec:backward}) to the process of identifying $S$ given the $O$ and
R. The Relation Coherence model is trained to compute a score of
each passage-answer candidate (see Section~\ref{subsec:coherence}). We call this
score coherence score because it indicates how compatible $S$, $R$, $O$ are
with each other in a passage. Finally,
each candidate-answer pair have two scores, the one provided by OpenQA which
constructs the prior and the coherence score computed by the Relation Coherence
model. Both scores are combined to get a final score for prediction as explained
in Section~\ref{subsec:ensemble}.


\mypar{Assumptions on Forward Searching} In addition to the candidate
passage-answers pairs, we assume Forward Searching has an encoding function
$fs\text{-}enc$ which returns a vector for each token in the question and
passage. Every modern OpenQA system has such an encoding function available.

\subsection{Backward Searching}
\label{subsec:backward}

Forward searching reduces finding the object in ($S_i$, $R_i$,
$\text{<}O\text{>}$) to a question answering problem. Inspired by this,
Backward searching also reduces the task of finding ($\text{<}S\text{>}$,
$R_i$, $O$) to a question answering problem. To solve this task, Backward
searching uses a Backward Reader, analogous to the Forward
Reader used by OpenQA. 

\mypar{Reverse Question} The backward searching process needs to answer the
reverse question to find the subject ($S$) in a candidate passage.  Constructing
the reverse question automatically is challenging.  We illustrate the
challenges with an example. To find the missing object in (USA, Capital City,
?), a forward question is ``What's the capital city of USA?''. Suppose that after
asking such a question OpenQA answers with: ``Washington, D.C.''. A reverse
question that given (?, Capital City, Washington D.C.) returns the subject is:
``Washington, D.C.  is the capital city of what country?''. The first challenge
of constructing such a reverse question is that it requires the name
of the target answer attribute, in this case ``country''. The second challenge
is determining the right interrogative word, e.g.  ``who'' or ``whose'' if the
subject is a person, and ``when'' if the subject is a date, etc. 


In our design we first define a new format for the reverse question and then
train a backward reader to understand this new format. Specifically, we
construct the reverse question by replacing the subject in the forward question
with a special token ``<sub\_mask>'', and then use two tags ``object :'',
``question : '' and a comma to combine the answer and the masked question. For
example, the reverse question would be ``$\textbf{object : } $ Washington, D.C.
\textbf{ , question :} What's the capital city of <sub\_mask>?''. Given this
reverse question and the subject as training examples, the Backward Reader
learns to find the correct subject for a new reverse question and passage.

The goal of the Backward Reader is to find the subject in a passage, so that we can
measure the coherence of the passage. To differentiate this subject from the
subject given in the
original question, we call the subject found by Backward Reader P-subject. The Backward Reader
is based on a pre-trained language model (we use BERT). Training the Backward
Reader involves fine-tuning the underlying pretrained BERT model which encodes
each token in the question and passage into a vector. To differentiate, we
denote by $bs\text{-}enc$ the encoding function in Backward searching.

\subsection{Coherence Score}
\label{subsec:coherence}

Coherence measures the total confidence in a passage of finding the object and
then back to subject . To this end, we add a back layer on top of OpenQA.
OpenQA compute the forward confidence score while the back layer computes the
backward confidence score. We train OpenQA and the back layer together to
maximize the coherence of correct passages. 

To measure the  backward confidence, we compute two scores, 

\mypar{One-hop subject similarity}  This measures how well P-subject matches the
subject mask ``<sub\_mask>'' in the reverse question.

\mypar{Two-hop subject similarity}  This measures how well P-subject matches the
subject in the forward question.

We construct each vector representation for subject,  <sub\_mask> and P-subject
and then use the approach in \cite{rank-fun-openqa, tree-based-conv} to measure
the similarity between two vectors. Specifically, given two vectors
$\textbf{h}_a$ and $\textbf{h}_b$, we measure the similarity as,
\[
sim(\textbf{h}_a, \textbf{h}_b, \boldsymbol{\theta}) = MLP_{\boldsymbol{\theta}}([\textbf{h}_a]; \textbf{h}_b; \textbf{h}_a \otimes \textbf{h}_b])
\]

Where $\otimes$ is the element-wise product operator, ";" is the column
concatenation operator, and MLP (multilayer perceptron) is
defined as $MLP(\textbf{h})=\textbf{w}_2^T(dropout(relu(\textbf{w}_1^T
\textbf{h} + \textbf{b}_1))) + \textbf{b}_2$, where $\textbf{w}_1, \textbf{w}_2,
\textbf{b}_1, \textbf{b}_2$ are network parameters, $relu$
\cite{Goodfellow-et-al-2016} is the activation function, $dropout$
\cite{Goodfellow-et-al-2016} is used to prevent overfitting. We use
$\boldsymbol{\theta}$ to denote the parameters in MLP.

We now show how to construct the  vector representation for them.

\mypar{Construct vector representation}  Backward Searching  takes answers
(including answer text and span) from OpenQA and constructs the reverse
questions, and feeds them to the backward reader to get P-subject.  This  shows
P-subject representation should be conditioned on the object representation which
has two versions, one from the Forward Searching  and other from Backward
Searching. We so use a MLP to learn a unified object representation. 

Denote by \textbf{f-span-h}(X) as the representation of a text span X from Forward
Searching, and ${\textbf{b-span-h}(Y)}$ as the representation of a text span Y
from Backward reader,  start(X) and end(X) as the start and end position of a
span X .  Specifically, \textbf{f-span-h}(X) = [$fs\text{-}enc$(X)[start(X)] ;
$fs\text{-}enc$(X)[end(X)]], which is the concatenation of first and last token
hidden state from the encoder function.  Then we have, 
\[
\textbf{h}_{uni\text{-}obj} = MLP_{\boldsymbol{\theta}}([\textbf{f-span-h}(object) ; \textbf{b-span-h}(object)]) 
\]
The P-subject representation used for similarity computation is thus,
\[
\textbf{h}_{p\text{-}sub} = [\textbf{h}_{uni\text{-}obj} ; \textbf{b-span-h}(P\text{-}subject)]
\]

 We also concatenate the unified object representation to that of subject and
<sub\_mask> for similarity computation,  because if P-subject is conditioned on
the unified object representation, the similarity should depend on it as well.
Another consideration is we must have all the vectors be the same dimension. So we
have
 \[
\textbf{h}_{sub} = [\textbf{h}_{uni\text{-}obj} ; \textbf{f-span-h}(subject)]
\]
 \[
\textbf{h}_{sub\text{-}mask} = [\textbf{h}_{uni\text{-}obj} ; \textbf{b-span-h}(sub\_mask)]
\]

We then compute the coherence score as:
\begin{alignat*}{2}
&S\text{-}Coher(Q,P,A) && = forward\text{-}score(Q,P,A, \boldsymbol{\theta}_1) + \\
&  &&1\text{-}hop\text{-}subject\text{-}similarity(Q,P,A, \boldsymbol{\theta}_2) + \\
&  &&2\text{-}hop\text{-}subject\text{-}similarity(Q,P,A, \boldsymbol{\theta}_2) \\
& 				  &&=forward\text{-}score(Q,P,A,  \boldsymbol{\theta}_1)  + \\
& &&sim(\textbf{h}_{p\text{-}sub}, \textbf{h}_{sub\text{-}mask} , \boldsymbol{\theta}_2) + \\
& &&sim(\textbf{h}_{p\text{-}sub}, \textbf{h}_{sub} , \boldsymbol{\theta}_2)
\end{alignat*}

We use the same $\boldsymbol{\theta}_2$ to share parameters for the one and two
hop subject similarity. Note the $forward\text{-}score$ contains new parameters
which means we retrain the OpenQA with backward searching to get the loop score.
This $forward\text{-}score$ is different from the OpenQA score used for
prediction which is trained independently from the Coherence Model.

\mypar{Offline training} We use cross-entropy loss \cite{Goodfellow-et-al-2016}
to train the coherence score model. Since it depends on OpenQA's output
passage-answers, we run OpenQA on training questions and retrieve 150 passages
from the corpus. We then pick the top M (we use 7) output passages for each
question as the training passages for the coherence model. If the top M output
does not contain positive or negative passages, we continue to search the
remaining output. If there is still no positive or negative passage we ignore
the training question. We make sure there are at least 2 positive/negative
passages for sample balance. We also need to train the backward reader and leave
the details in Section~\ref{sec:train_back_reader}.

\subsection{OpenQA and Coherence Ensemble}
\label{subsec:ensemble}

The Relation Coherence model depends on the top passage-answers from
OpenQA which already assigns a score to each. We combine OpenQA's and
Relation Coherence's scores into the final score. 

\mypar{Ensemble of experts} We have two experts, OpenQA and Relation
Coherence, each of which gives a different score to a $(Q, P, A)$ tuple.  A
naive combination of these scores, for example via summation or average, does
not yield good results because these scores were computed by different models
and their ranges differ.  Instead, given a relationship and an expert, we use
Z-score ~\cite{feature_scaling} to normalize the expert score of all tuples in
that relationship and we call it across-row-normalization. This strategy
exploits information across rows of the same relationship. 

\mypar{FabricQA-Extractor Inference} After normalizing scores, we add
them to obtain the final prediction for tuple
$(Q, R, A)$, \ie
\begin{alignat*}{2}
&(P,A)^* = \argmax_{(P,A)}(&&z\text{-}tn(S\text{-}OpenQA(Q,P,A)) + \\
&         				   &&z\text{-}tn(S\text{-}Coher(Q,P,A)))
\end{alignat*}

where $z\text{-}tn(...)$ is the Z-score table normalization function.

The Relation Coherence model exploits the relational information to
boost the quality of the extraction. It depends on an OpenQA system to obtain a
candidate of passage-answer pairs $(P, A)$. We now turn our attention to the
end-to-end system design, FabricQA-Extractor.

\section{FabricQA-Extractor Architecture}
\label{sec:fullsystem}

In this section, we present FabricQA-Extractor's overview in 
Section~\ref{subsec:overview}. We explain the query lifecycle in
Section~\ref{subsec:lifecycle} and important implementation details in
Section~\ref{subsec:offline}.

%

\subsection{System Overview and Design Goals}
\label{subsec:overview}

\begin{figure}
  \centering
  \includegraphics[width=\columnwidth]{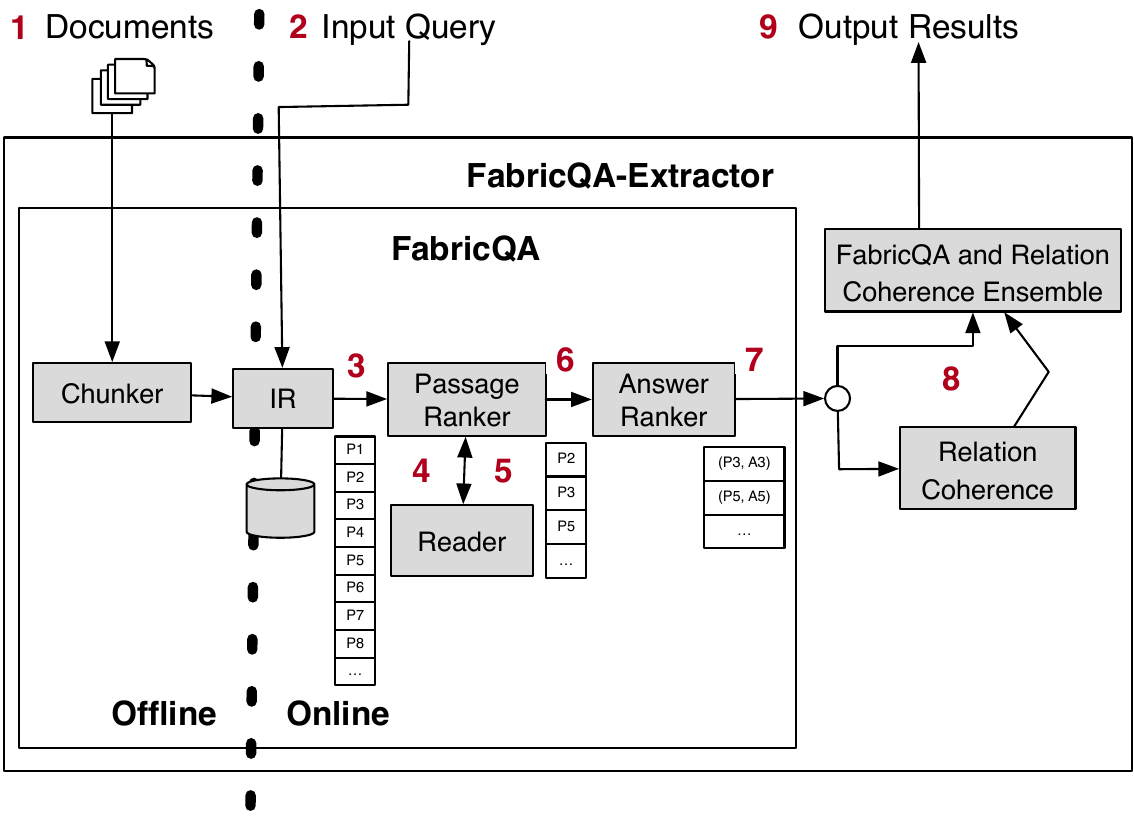}
  \caption{FabricQA-Extractor's architecture overview}
  \label{fig:architecture}
\end{figure}

The FabricQA-Extractor system architecture is shown in
\F\ref{fig:architecture}.  This architecture is the result of several
iterations. 

FabricQA-Extractor takes as input a partially filled relation and a
collection of documents. It then finds a set of suitable passages, and for each
passage, a span of text that answers the question. The question answering
component, named FabricQA in \F\ref{fig:architecture}, returns 
top-$K$ answers to the Relation Coherence model.

FabricQA-Extractor implements a funnel-based architecture. The
more downstream a component, the more sophisticated and the more computationally
expensive it is. The system is designed so that as data flows downstream, it is
increasingly filtered out by each component. As a result, the system achieves
high processing throughput. 

\mypar{Design Goals} There are 4 metrics that drive the design of this system.
\emph{Accuracy:} we design the system to achieve good extraction quality,
measured as the number of correct answers. \emph{Scalability:} The system must
work even on tens of millions of documents. \emph{Throughput:} We want to answer
questions with subsecond latencies to make the system practical.
\emph{Transparency:} The system consists of a complex pipeline of components. We
strive to attach context (document, passage, answer, scores) to each provided
answer to facilitate the system's use as explained in the previous section.

\subsection{Query Lifecycle}
\label{subsec:lifecycle}

We explain the online and offline components: 

\subsubsection{Chunker and Indexer (Offline)} 
\label{subsec:chunkerindexer}

During an offline stage, the collection of documents is indexed into an
information retrieval system\footnote{We use
ElasticSearch~\cite{elasticsearch}}. The indexing strategy consists of splitting
documents into passages of roughly the same size and often consists of
several sentences. This size is compatible with state of the art RC models,
which contributes to extracting good quality from those models.  For example,
BERT-based QA readers such as \cite{albert} do not accept passages longer than
512 subword tokens, and a single word is usually represented with more than one
subword token, according to the wordpiece algorithm~\cite{wordpiece}.
The released BERT-based reader implementation
\cite{google_bert_code} uses at most 384 subword tokens and we split passages
into 100 words approximately and show evidence for this decision in the
evaluation section.

The chunker component of FabricQA reads documents as a stream
of tokens. To chunk the stream into passages, it uses a sliding window that
ensures there is some overlap of tokens across neighboring paragraphs. Although
this redundancy increases the storage footprint in the information retrieval
system, it also boosts the performance of the end to end system because it
avoids possible context loss.

Finally, during the offline stage, the Indexer component assigns an ID to each
passage and stores it along with the document it belongs to in the information
retrieval system.

\mypar{Note on information retrieval system} We assume the information retrieval
system can return the $K1$ most relevant passages to an input query, along with
the ranking score associated with each passage. Most information retrieval systems
implement this functionality. 

\subsubsection{Passage Ranker (Online)}
\label{subsec:passageranker}

The Passage Ranker retrieves $K1$ passages from the information
retrieval system and forwards $K2$ passages downstream to the Answer
Ranker, where $K2 < K1$.  $K1$ is usually in the hundreds, while $K2$ in the
tens. The information retrieval system can return those passages fast because it
only uses \emph{syntactic} information to rank and retrieve passages. The
Passage Ranker uses \emph{semantic} information, as available through
pre-trained language models, to narrow down the passages that get sent
downstream to the more computationally expensive Answer Ranker.

After extensive experimentation, we concluded that modern RC models subsume the
role of a Passage Ranker. That is, 
the scores they associate to the extracted answers already rank the
passages according to how likely they are to contain the right answer to the
question. We exploit this insight to implement the Passage Ranker as
explained next.

\mypar{RC Model} An RC model takes as input a question and a passage and finds a
span of text in the passage that corresponds to the answer. It then returns the
text span and a score, $S_{best}$ associated with the confidence the model has in
that answer. Furthermore, state-of-the-art RC models return ``null'' when they do
not find a valid text span in a passage \cite{bert}. 

\mypar{Building a Passage Ranker off RC Model scores} We apply the RC model to
each passage obtained from the information retrieval system. Then, we normalize
the scores associated with each passage-answer using a simple \emph{margin
score}, defined as: $S_{margin} = S_{best} - S_{null}$. Then, this component
sends downstream the top-$K2$ passage-answer-question triples, those with the
highest $S_{margin}$. 

\subsubsection{Answer Ranker}
\label{subsec:answerranker}

The Answer Ranker is a neural network model that takes a question, $Q$,
passage $P$, and the answer provided by the Reader $A$ and returns a score that
indicates how likely $A$ is an answer to $Q$ given $P$. Each of the $K2$
passages from the Passage Ranker is fed to the Answer Ranker
and the top-$K3$ chosen. Note one can configure $K3=1$ to obtain the
best answer according to FabricQA. In general,
FabricQA-Extractor will obtain $K3>1$ and provide those candidate
answers to the Relation Coherence model.

\begin{figure}[h]
  \centering
  \includegraphics[width=\columnwidth]{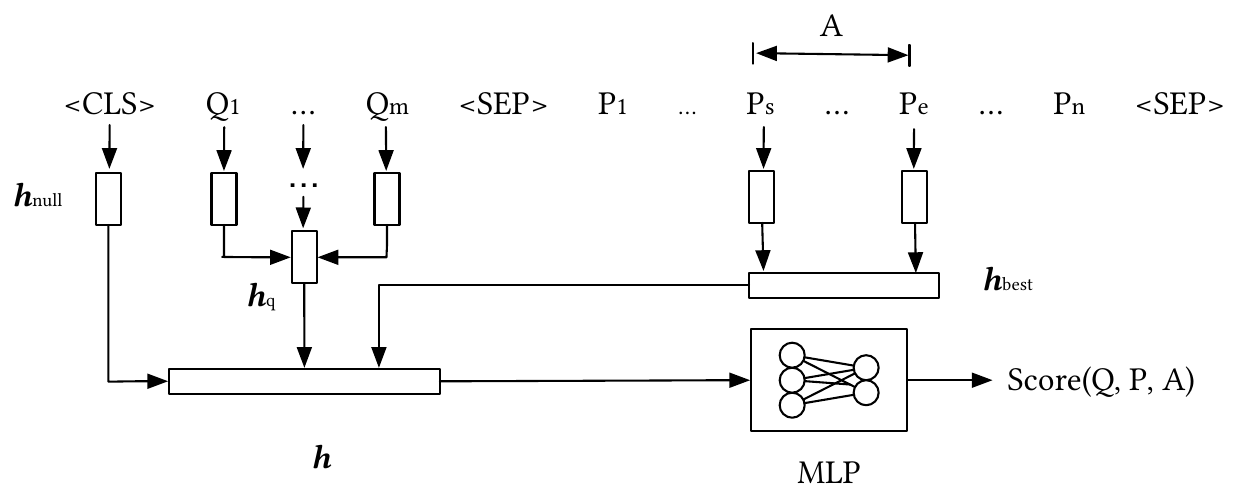}
  \caption{Schematic of the Answer ranker model}
  \label{fig:answer_ranker}
\end{figure}

$P$, $Q$, and $A$ are featurized into fixed-length vectors using BERT as the
language model. The resulting feature vectors are concatenated as shown in
\F\ref{fig:answer_ranker}.  $Q$ and $P$ are prefixed by a special token $\text{<}CLS\text{>}$
and there is a special end token, $\text{<}SEP\text{>}$, after which the answer is included.
The answer, $A$ is represented as a continuous span of tokens from $P_s$ to
$P_e$, which correspond to the start and end tokens indicating the span of text
in the passage that corresponds to the answer.

The model uses the following vector representations:

\begin{myitemize}

\item{$\textbf{h}_{best}$}, vector representation for the answer. We concatenate
the answer start and end token vector along column to get the best answer
vector, i.e. $\textbf{h}_{best} = [BERT(P_s);BERT(P_e)]$. Where
$\text{BERT}(P_i)$ denotes the vector of token $P_i$ encoded by Bert.

\item{$\textbf{h}_{null}$}, is the vector representation for cases where no
answer is found, in which BERT returns the position of the start token, $\text{<}CLS\text{>}$. 

\item{$\textbf{h}_{q}$}, is the vector representation of the question, computed
as the average of the vector representation of each token in $Q$:
$\textbf{h}_{q} = \frac{1}{m}\sum_{i=1}^{m}\text{BERT}(Q_i)$

\end{myitemize}

The model then obtains \textbf{h} by concatenating the previous vectors, $\textbf{h} =
[\textbf{h}_{best} ; \textbf{h}_{null} ; \textbf{h}_{q}]$, and then feeds it
into a multilayer perceptron to compute the score: $S(Q,P,A) = MLP(\textbf{h})$.

The Answer Ranker models the distance between the answer to the
question and to the null answer in an embedded space, $S(Q, P, A)$. Scores are
assigned based on the distance to question and null answer.

\subsection{Implementation Details}
\label{subsec:offline}

We provide a list of important implementation details 
that proved critical to the end to end performance of
\textsf{FabricQA-Extractor}.

\mypar{Answer Ranker Offline training} To train
Answer Ranker, for each Q, we need $N^+$ positive passages $P^+$ and $N^-$
negative passages $P^-$. We use cross-entropy loss as the objective function.

The training dataset for the Answer
Ranker consists of tuples of the form (Subject, Relationship, Question template,
Object, Positive Passages, Negative Passages). We collect the passages using
distant supervision \cite{distantsupervision}. To do that, we expand the question template using the
subject, and then use the resulting question to identify top-150 relevant
passages using the IR system. We then use the reader to obtain an answer from
each passage. Finally, we compare the answer to the object using fuzzy matching.
When the $F1$ score between answer and object is above a $0.7$ threshold we
label the passage positive, and we label it negative otherwise. We ensure that
each tuple in the training dataset has at least 1 positive passage and 5
negative ones. This automatic approach makes it practical to assemble a training
dataset to fit the models.

\mypar{Building training datasets for Backward Reader}
\label{sec:train_back_reader}
To train the Backward Reader of the Relation Coherence model, we follow an automatic approach. 
The gist of the method is to invert subject and object. 
The trick we implement is to invert the question, Q. To do that, we
replace interrogative adverbs like $what$, $when$, $who$ with the subject, and
the subject with all possible interrogative adverbs to obtain a list of reverse
questions.  For example, given a question template \emph{"What organization did
<person> work for?"} and the subject (answer) \emph{"IBM"}, and given a passage,
we produce the following ungrammatical questions: \emph{"IBM organization did
who work for?"}, \emph{"IBM organization did what work for?"}, etc. We then feed
each reverse question and the passage to the Forward Reader. If any answer
produced by the Forward Reader is close to the subject (fuzzy matching), we keep
the passage as the training data and also label the answer span as P-subject. 
Despite the noisy process, this approach is effective and helped us
train the Backward Reader and thus coherence score model to boost the performance of the entire
system.

\mypar{Question pre-processing} Before sending a question to the information
retrieval system, the question is
pre-processed with Spacy~\cite{spacy2} to identify important tokens, such as the
Saxon genitive, which indicates possession and has thus an important impact on
the meaning of questions. For example, default tokenizers will produce the token
'John's'; with the pre-processing we seek to have 'John' and ''s' as separate
tokens. 

\mypar{Passage Re-ranking} Given two passages that contain the tokens in
the subject, we would like the passage matching exactly the subject phrase to
have a higher score. However, ElasticSearch uses BM25 \cite{bm25, INR-019} to
score passages given a question and BM25 does not consider token order. We,
therefore, construct a compound bool query to ensure exact match (including
order) ranks higher. 

\mypar{Exposing model internal representations} Commoditization of machine
learning models means that many models are published as black boxes, which is
good for usability, but insufficient when access to their internal
representations is needed. We modify the RC model available on Hugging
Face~\cite{hugging_face} so we access its internal token vector representation,
which we use for training the Relation Coherence model.

\section{Evaluation}
\label{sec:evaluation}

In this section, we answer the main research questions of our work:

\begin{myitemize}


\item \textbf{RQ1. Does the Relation-Coherence model improve the extraction
quality?}

\item \textbf{RQ2. Does the Relation-Coherence model improve the extraction
quality \emph{without} retraining?}

\item \textbf{RQ3. Is FabricQA-Extractor fast enough to be used in
practice?}

\item \textbf{RQ4. How sensitive is the Relation-Coherence model to the training
dataset?}

\end{myitemize}

We also include a microbenchmark section to study other relevant aspects of the
system. In particular:

\begin{myitemize}

\item \textbf{RQ5: Is FabricQA a strong baseline?}

\item \textbf{RQ6: How does splitting passages affect quality?}

\end{myitemize}

We first describe the metrics and the setup we use in the experiments. We
explain the datasets in context with the relevant experiments. 

\subsubsection*{\textbf{Metrics}} To measure the extraction quality we use
several standard metrics often used in the literature:

\noindent\underline{Exact Match (EM)}: For each question and answer pair, this
metric takes as value 1 when the answer exactly matches the ground truth answer,
and 0 when it does not. Both ground truth and the provided answer are first
normalized (removing articles, punctuation, convert characters to lower case)
using the same procedure. 

\noindent\underline{F1 Score (F1)}: Because the EM metric is sensitive to the
specific tokens in the answer (\eg "the 4th of July" vs "July, 4th"), the F1
score is computed at the token level, using the precision (\#correct tokens in
answer / tokens in predicted answer) and recall (\#correct tokens in answer /
tokens in ground truth answer). 

\noindent\underline{Aggregated EM/F1}: In general, we are interested in
understanding the relative performance of a system on tables that contain
multiple questions. We aggregate the metrics of the questions within
each table and report that result.

\subsubsection*{\textbf{System Setup}} We run all the experiments on Chameleon
Cloud \cite {keahey2020lessons}. We use a single node which has two Intel Xeon
CPUs, each with clock speed of 2.60 GHz. In total the node has 48 threads and
187G RAM. In addition, it has one NVIDIA RTX 6000 GPU on which most of the system
computation runs. The operating system is Ubuntu 18.08 and the CUDA version is
10.1. The system is implemented in Python v3.7.9.

\subsection{RQ1: Does the Relation Coherence model help improve extraction
quality?}

In this section, we evaluate the contribution of the Relation Coherence model to
the end-to-end problem of filling in incomplete tables from large collections of
documents. We use the Relation Coherence model as part of our new FabricQA-Extractor
system and as as part of DrQA~\cite{drqa}, an open-source state of the art open question
answering system. 
We first introduce datasets, baselines, and then present the results.



\mypar{Datasets} We use the QA-ZRE relation extraction
benchmark~\cite{zero-shot-rel}.  In this benchmark, the goal is to find the
answer to a question \emph{given a passage}. This is the benchmark used to
evaluate the Zero-Shot model~\cite{zero-shot-rel}. The benchmark includes 120
relationships and 1192 questions. The statistics of the dataset are summarized
in Table~\ref{tab:qa_zre_data_summary}. 

\begin{table}[h]
\center
\begin{tabular}{|l|l|r|}
\hline
Relationships                                                                                   & \multicolumn{2}{r|}{120} \\ \hline
\multirow{4}{*}{\begin{tabular}[c]{@{}l@{}}Tuples(Rows)\\ Per Relationship\end{tabular}}        & Min         & 172        \\ \cline{2-3} 
                                                                                                & Max         & 5,274      \\ \cline{2-3} 
                                                                                                & Average     & 3,213      \\ \cline{2-3} 
                                                                                                & Total       & 385,610    \\ \hline
\multirow{4}{*}{\begin{tabular}[c]{@{}l@{}}Question Templates \\ Per Relationship\end{tabular}} & Min         & 1          \\ \cline{2-3} 
                                                                                                & Max         & 35         \\ \cline{2-3} 
                                                                                                & Average     & 10         \\ \cline{2-3} 
                                                                                                & Total       & 1,192      \\ \hline
\end{tabular}
\caption{QA-ZRE dataset data statistics}
\label{tab:qa_zre_data_summary}
\end{table}

Each relational tuple contains the subject and relationship but lacks the
object. Along with each tuple, the benchmark includes a \emph{passage} on which
to find the object.  Instead of using the benchmark directly, we solve the
harder problem of first finding the passage among all passages in the input
document corpus and then finding the answer in that passage. We use a dump of
the English \textsc{Wikipedia} which contains 5,075,182 documents that leads to
35,338,951 passages: the goal is to find an answer to a question in a passage
among 35M passages. This harder setup corresponds to the problem we target in
this paper.


\mypar{Baselines} We measure the extraction quality of
FabricQA-Extractor against other strong baselines.

\noindent$\bullet$~FabricQA. This is FabricQA-Extractor but
without using the Relation Coherence model. This benchmark sheds light
on the marginal contribution of the Relation Coherence model to the
extraction quality.

\noindent$\bullet$~FabricQA-Ensemble.
We combine any two model instances $Model_i$ and $Model_j$ of Answer Ranker
into a new model $Model_{(i,j)}$ by averaging their prediction scores. We choose
the best $Model_{(i,j)}$ compared to the baseline Answer Ranker instance on
development data and call it FabricQA-Ensemble. This simple ensemble technique is popular in 
machine learning community and more robust than a single model \cite{pml1Book, dong2020survey}.

\subsubsection{Relation Coherence Evaluation on FabricQA-Extractor}
\label{subsubsec:mainresults}

\mypar{Experimental Setup} We split the dataset 70/10/20 into train, dev, test
splits, leading to a total of 24 partially filled tables that contain subjects
but have 70,834 missing cells that correspond to the objects. We report the
extraction values based on the test data.  To conduct the experiment, we feed
each partially filled relation to the system, that fills in the relation by
searching over the 35M passages of the English Wikipedia. We configure
FabricQA-Extractor as follows: the Passage Ranker retrieves 30
passages from the IR system and FabricQA forwards the top 5 passages
for each question to the Relation Coherence model, which selects the
final answer.

\mypar{Results} We first use FabricQA (\ie without Relation
Coherence) to fill in the tables by choosing the top-1 answer and measure
its aggregated EM and F1 performance. We use its performance as the
baseline for this experiment. 

\begin{figure}[t]
  \centering
  \includegraphics[width=\columnwidth]{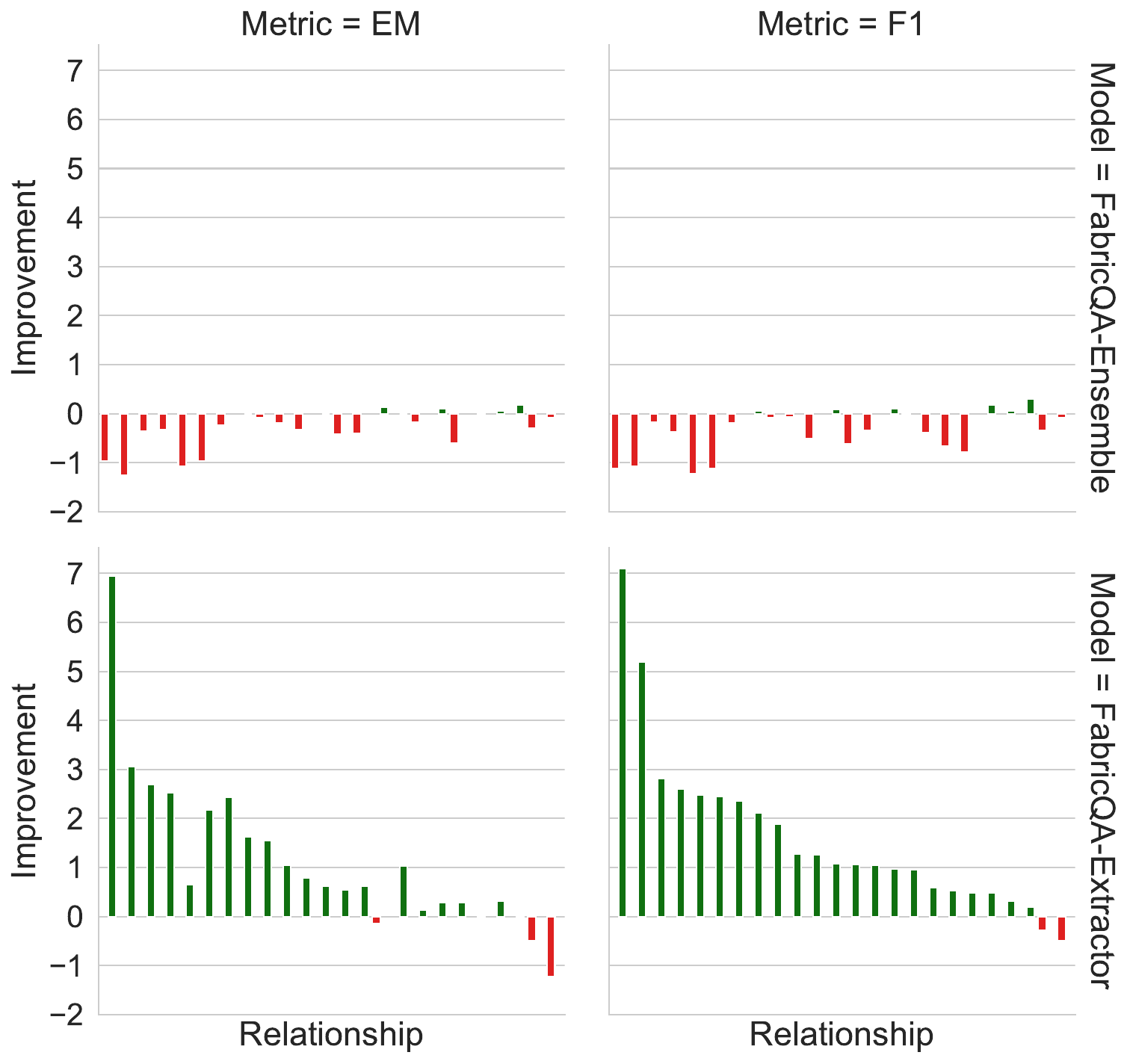}
  \caption{FabricQA-Ensemble relative improvement over
FabricQA (top). FabricQA-Extractor improvement is shown at the
bottom.}
  \label{fig:qa_zre_model_improvement_dist}
\end{figure}

\F\ref{fig:qa_zre_model_improvement_dist} shows the relative performance of both
baselines,  FabricQA-Ensemble, and FabricQA-Extractor.  The $x$
axis shows all 24 tables on which we evaluate the systems. The $y$ axis shows
the relative improvement (or degradation) of performance using both the
aggregated EM (left column) and F1 metric (right column).
FabricQA-Ensemble yields worse results for $66.7\%$ of the
relationships. 

In contrast, FabricQA-Extractor improves the results for $91.7\%$ of
the tables (F1) and $87.5\%$ (EM). This demonstrates the effectiveness of
exploiting relational knowledge for information extraction via the
Relation Coherence model.

The above results are obtained when the Relation Coherence model was
trained with 70\% of the dataset. We now turn our attention to understanding in
depth what is the effect of the training dataset on the extraction quality.

\subsubsection{Relation Coherence Evaluation On DrQA}

\mypar{Experimental Setup} We first train DrQA \cite{drqa} using the same training data used
for FabricQA to isolate performance differences to the effect of the
Relation-Coherence model. We call the resulting system DrQA-Adapted.

\begin{figure}[t]
  \centering
  \includegraphics[width=\columnwidth]{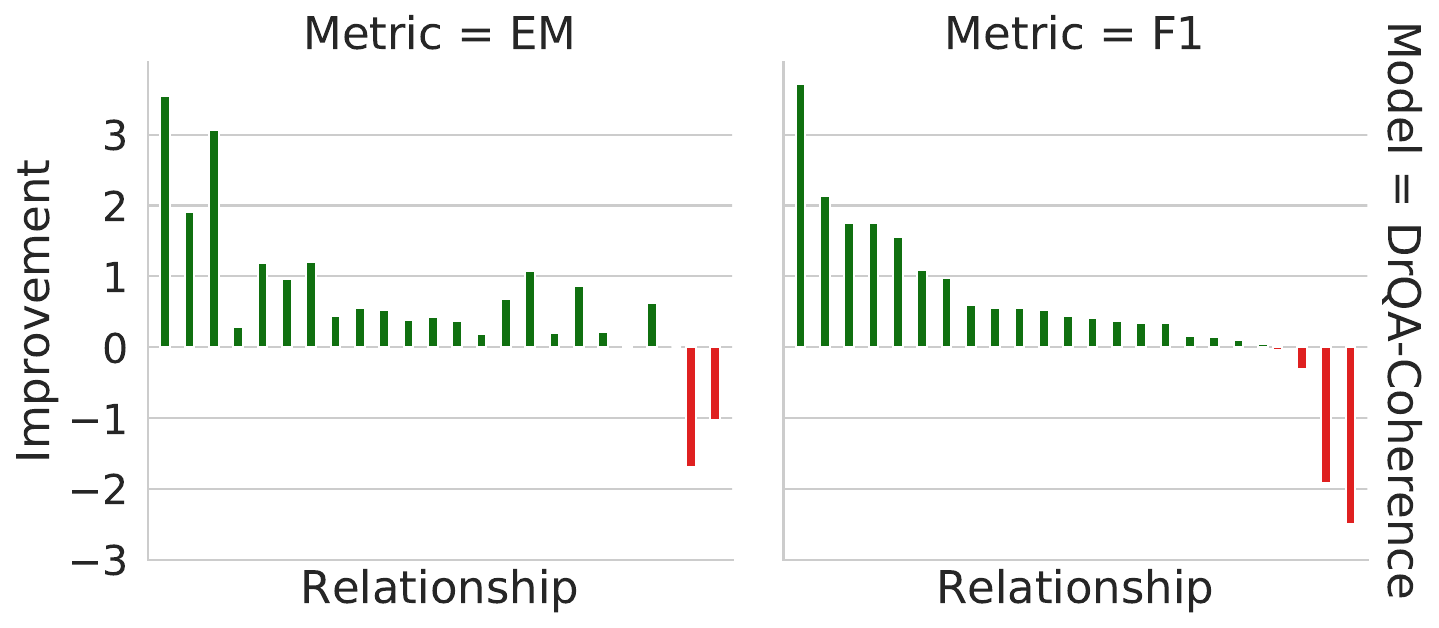}
  \caption{DrQA-Coherence relative improvement over DrQA-Adapted}
  \label{fig:drqa_coherence_improvement_dist}
\end{figure}

\mypar{Results} \F\ref{fig:drqa_coherence_improvement_dist} shows the relative
performance of DrQA-Coherence w.r.t DrQA-Adapted. The results show that
the Relation-Coherence model improves DrQA-Adapted on
$83.3\%$ of the relationships, both in EM and F1. The improvement magnitude is
not as high as in FabricQA-extractor. This is because the DrQA reader performs worse than the reader used
in FabricQA-extractor.

This result demonstrates the ability of Relation-Coherence to use the target
relation to improve the extraction quality of OpenQA systems.


\subsection{\textsf{RQ2: Does Relation Coherence improve extraction quality
without retraining?}}

In this section, we measure the Relation Coherence extraction quality
\emph{on a different dataset}, with new relationships, and without retraining. 

\mypar{Datasets and Baselines} We use the BioNLP benchmark used as part
of the BioNLP Open Shared Tasks 2019 (BioNLP-OST-2019)~\cite{bionlp-ost-2019}.
This is a benchmark for relation extraction in the biomedical domain. Similar to
QA-ZRE, we use this benchmark to derive a series of partially filled relations.
While the benchmark provides the passages from where to extract the answers, we
instead solve the harder problem of identifying those passages in a document
corpus. We use the \textbf{PubMed} corpus that contains 24,612,524 abstracts in
biology, botany, and biomedicine domain which results in 81,641,516 passages
when indexed in FabricQA-Extractor. 
So we need to
find the right answer among the 81M.  We use the same baselines as in the
previous section.

\mypar{Experimental Setup} We measure the performance of
FabricQA-Extractor on the BioNLP benchmark when finding the answers
among the 84M passages from PubMed without retraining the system. We do that to
understand how FabricQA-Extractor performs on a different dataset and,
more importantly, to determine whether once trained (in this case on Wikipedia
data), the Relation Coherence model can use relational information in
other domains.

Unlike with the previous dataset, we do not have ground truth answers
for the BioNLP questions over PubMed data. We hire a domain
expert to manually label the question-answer pairs as correct and incorrect.
Then, instead of using EM and F1 as metrics, we use the correct ratio of answers
for each relationship.

Due to the cost of manual labeling we use 10 relations in this experiment and
30 rows per relation, totaling 300 questions.  We write a question template for
each relationship and obtain the top passages (and answers) by FabricQA
and FabricQA-Extractor.  We ask the expert to read each question and
passage to determine whether the predicted answer is correct. In total, the
expert manually labels 600 passage-answer pairs. 

\mypar{Results} \F \ref{fig:general_throughput_effect_passage} a) shows the extraction quality of FabricQA and
FabricQA-Extractor on the sample BioNLP data.  Without retraining, the
Coherence Model improves the quality of FabricQA on a new domain on
60\% relationships with a large margin (> 8.5\%). 
When it
performs worse, it is with a small margin (< 3.5\%). In addition, both systems
tie on 10\% relationships. On average, Coherence Model improves the
quality by 7 points on this biomedical domain. 

Overall the result shows the capacity of the Relation Coherence model's
to exploit relational information in new domains without retraining.
We consider these results are the strongest evidence for the opportunity of
using relational information in information extraction tasks that use question
answering technology.

\subsection{RQ3: FabricQA-Extractor Throughput}

In this section, we demonstrate how the funnel architecture of
FabricQA-Extractor leads to high performance. Recall from
Section~\ref{subsubsec:rc} that applying an RC model to each passage
would lead to multi-day latencies to solve problems of the scale we show here.
Instead, FabricQA-Extractor answers questions in subsecond latencies.
We use different baselines to demonstrate the performance characteristics of
each pipeline component:

\mypar{Experimental Setup and Baselines} We use a sample of 100 relation triples
(questions) from the test data to conduct this experiment. We measure throughput
as \#questions answered per second.  For FabricQA and
FabricQA-Extractor, IR retrieves 30 passages for each question and
sends them to Answer Ranker, which picks top 5 answers and forwards
them to Relation Coherence model.

\noindent$\bullet$~IR. This baseline corresponds to the time to extract
the 30 passages from the IR system and preprocessing the questions as explained
in Section~\ref{sec:fullsystem}. This is measuring the performance of the
information retrieval system: no ML model inference is performed at this stage.

\noindent$\bullet$~IR + Reader This baseline corresponds to the
previous + the execution of the Passage Ranker, which requires calling
an RC model, the Reader over each of the 30 passages. The inference is
performed on the GPU.

\noindent$\bullet$~FabricQA. The previous plus the time to call the Answer Ranker, which ranks the final answers before forwarding them to
the \textsf{Relation Coherence} model.

\noindent$\bullet$~FabricQA-Extractor This is the whole pipeline.

\begin{figure*}[h]
    \centering
    \subfloat[Left. Performance on BioNLP-Open dataset]{{\includegraphics[width=0.7\columnwidth]{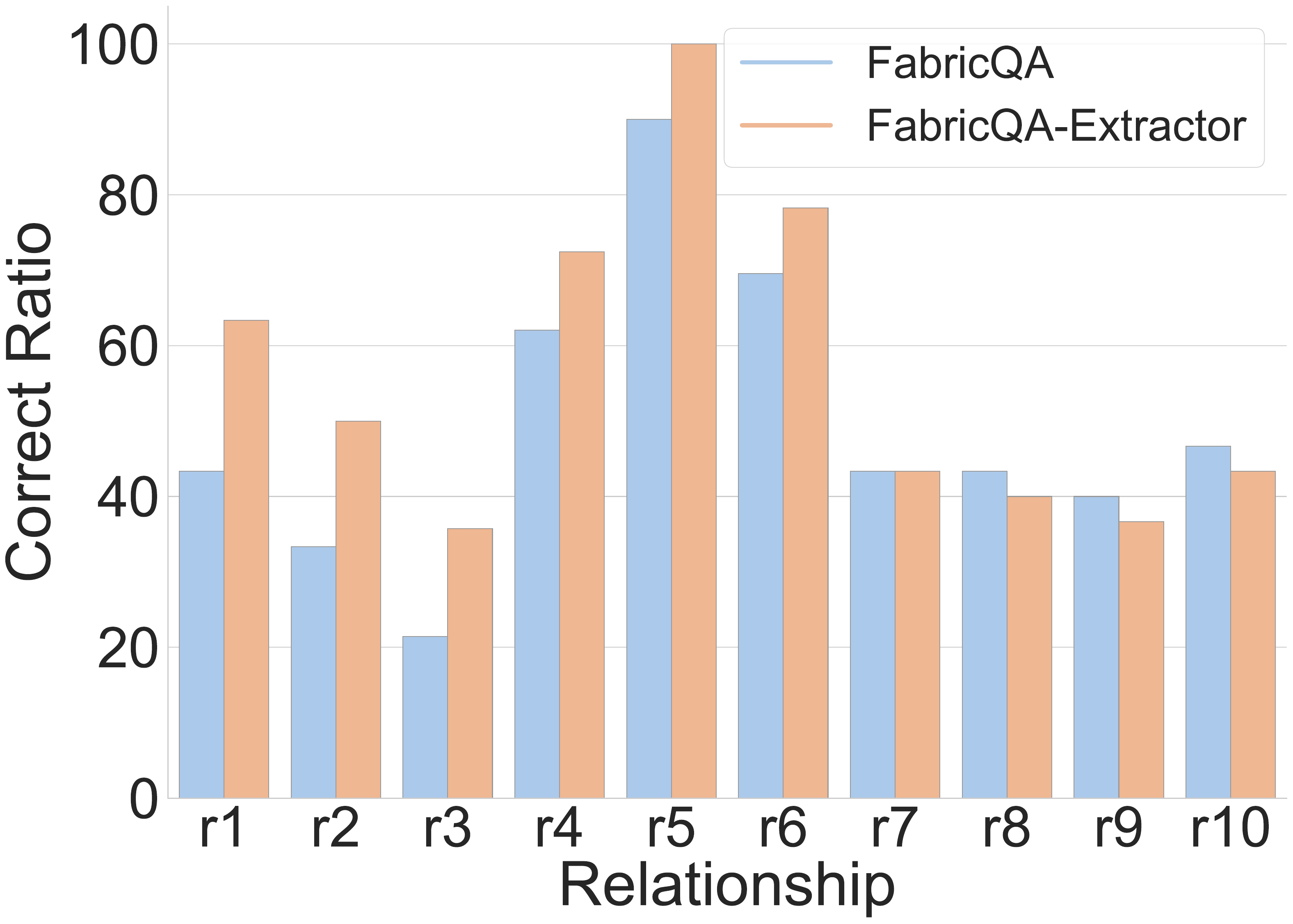}}}%
    \subfloat[Center. Pipeline Throughput]{{\includegraphics[width=0.7\columnwidth]{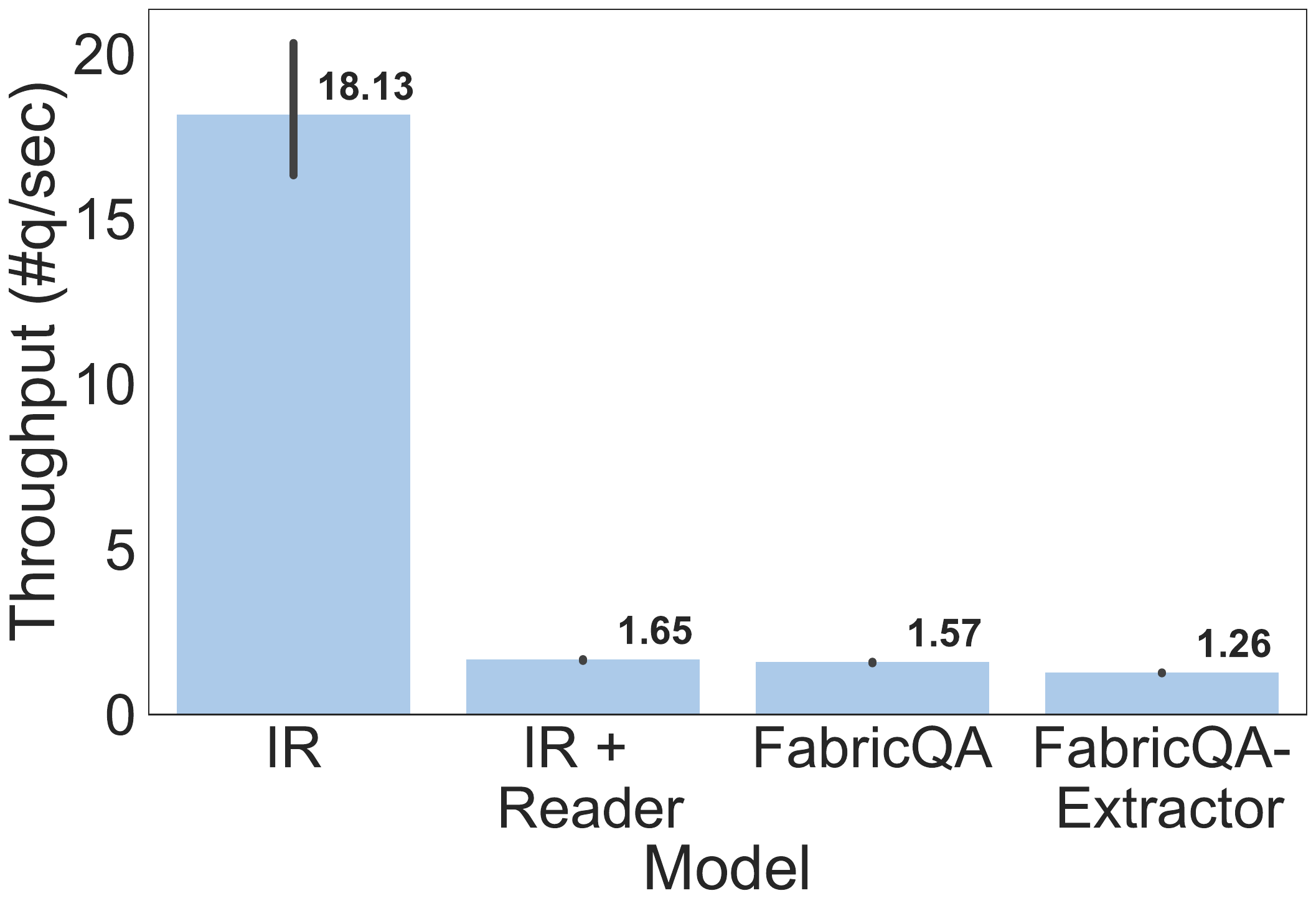} }}%
	\subfloat[Right. Extraction performance for different chunk sizes]{{\includegraphics[width=0.7\columnwidth]{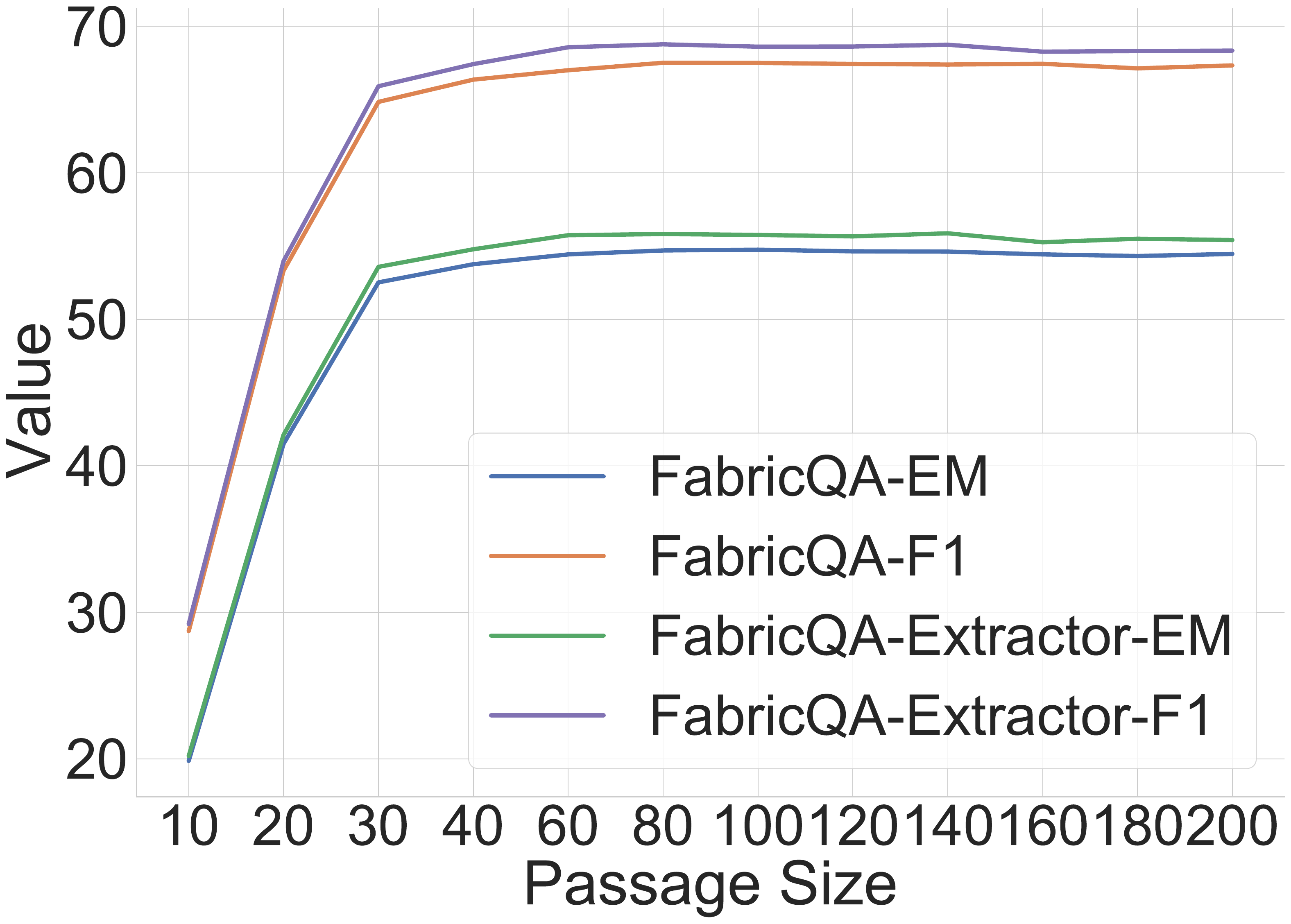}}}%
	\caption{Generalization experiment (left). Throughput experiment (center).
Effect of passage size (right).}
	\label{fig:general_throughput_effect_passage}
\end{figure*}

\mypar{Results} Figure \ref{fig:general_throughput_effect_passage} b) shows the throughput of the different components of the
pipeline. We highlight 3 results. First, and most obvious, as soon as deep
neural models are incorporated in the pipeline, the throughput drops with
respect to the IR component, despite this computation taking place on a
GPU. Second, the further downstream we go in the pipeline, the lower the
throughput, as components become more sophisticated.
Note how large is the performance factor between
baselines. For example, FabricQA is slower than IR + Reader
despite processing only 16\% of the data. The performance would drop fast if it
was not for the funnel design of our system, where each component vastly reduces
the workload for the downstream component. In conclusion,
FabricQA-Extractor provides subsecond latencies, making it practical to
address large scale extraction problems.

\subsection{RQ4: Sensitivity to Training Dataset Size}
\label{subsubsec:sampleefficiency}

We want to understand how sensitive the Relation Coherence model is to
the training dataset. Ideally, the model learns how to exploit the relational
information after seeing a few samples, hence yielding a boost in the extraction
quality with a small (and cheap) training dataset. 

To study this problem we use two different strategies to construct training
datasets of varying sizes. In the first strategy, we create a training dataset
by sampling tuples at random from each relationship. This
strategy, called \textbf{changing rows}, produces a training dataset that has
samples \emph{from every relationship}. In the second strategy, we fix the
number of tuples per relation and vary the number of relations from where we
sample. This strategy, called \textbf{changing relationships}, produces a
training dataset with \emph{samples from a subset of the relationships}. We also
report results when the model is not trained, as a reference baseline.

\mypar{Results} Table \ref{tab:row_efficiency} shows the result of the
\textbf{changing rows} strategy. In this case, the results correspond to the
aggregation of EM/F1 not only across rows of a table but across tables as well.
The results show that without any training data, FabricQA-Extractor
achieves only 3.11/6.94 (EM/F1). But then, using only 1\% of data (which
corresponds to 2208 rows) results in 55.27/68.46 the biggest jump in
performance. From then on, we observe two trends. First,
Relation Coherence strictly improves extraction quality compared to
FabricQA. Second, the results demonstrate the diminishing returns of
training data. The maximum extraction quality (56.24/69.3) is achieved when
using 25\% rows per relationship, which corresponds to 55200 rows, or 20x the
training dataset size for an improvement of less than 1 point. In other words, a
small, representative training dataset is sufficient for
Relation Coherence model to exploit relational data and improve the
extraction quality. 

\begin{table}[ht]
\begin{tabular}{|c|c|c|c|c|c|}
\hline
\multicolumn{2}{|c|}{\textbf{Rows(Questions)}} & \multicolumn{2}{c|}{\textbf{FabricQA}} & \multicolumn{2}{c|}{\textbf{FabricQA-Extractor}} \\ \hline
\textbf{Percent}        & \textbf{Count}       & \textbf{EM}        & \textbf{F1}       & \textbf{EM}             & \textbf{F1}            \\ \hline
0                       & 0                    & 4.29               & 8.02              & 3.11                    & 6.94                   \\ \hline
0.1                     & 235                  & 53.81              & 66.76             & 54.92                   & 68.23                  \\ \hline
0.5                     & 1,121                & 54.44              & 67.54             & 55.32                   & 68.64                  \\ \hline
5                       & 11,059               & 54.50              & 67.33             & 55.57                   & 68.73                  \\ \hline
15                      & 33,183               & 55.13              & 67.93             & 56.04                   & 69.3                   \\ \hline
25                      & 55,307               & 54.94              & 67.8              & 56.24                   & 69.3                   \\ \hline
\end{tabular}
\caption{Sample efficiency with changing rows}
\label{tab:row_efficiency}
\end{table}

To study the effect of training datasets that do not represent every
relationship in the test data, we use the \textbf{changing relationships}
sampling strategy. Table~\ref{tab:relation_efficiency} shows the results.
The trend is similar to the previous one. Sampling from only 5/24 relationships
leads to a quality of 52.52/65.58 while sampling from 20/24 only increases that
quality by approximately a point (EM). Consistent with the previous strategy,
Relation-Coherence improves the extraction quality when compared to
FabricQA. Overall, the quality achieved with the \textbf{changing
relationship} strategy is slightly inferior to the one achieved with
\textbf{changing rows}, which is expected because the training dataset is not as
representative as the one produced by the previous strategy.

\begin{table}[ht]
\begin{tabular}{|c|c|c|c|c|c|}
\hline
\multirow{2}{*}{\textbf{\begin{tabular}[c]{@{}c@{}}Relation-\\ ships\end{tabular}}} & \multirow{2}{*}{\textbf{\begin{tabular}[c]{@{}c@{}}Rows\\ (Questions)\end{tabular}}} & \multicolumn{2}{c|}{\textbf{FabricQA}} & \multicolumn{2}{c|}{\textbf{\begin{tabular}[c]{@{}c@{}}FabricQA-\\ Extractor\end{tabular}}} \\ \cline{3-6} 
                                                                                    &                                                                                      & \textbf{EM}        & \textbf{F1}       & \textbf{EM}                                  & \textbf{F1}                                  \\ \hline
0                                                                                   & 0                                                                                    & 4.29               & 8.02              & 3.11                                         & 6.94                                         \\ \hline
1                                                                                   & 98                                                                                   & 41.02              & 52.6              & 44.64                                        & 56.68                                        \\ \hline
5                                                                                   & 490                                                                                  & 51.18              & 64.23             & 52.52                                        & 65.58                                        \\ \hline
10                                                                                  & 980                                                                                  & 53.58              & 66.85             & 54.89                                        & 68.28                                        \\ \hline
20                                                                                  & 1,960                                                                                & 53.57              & 66.5              & 55.16                                        & 68.43                                        \\ \hline
40                                                                                  & 3,920                                                                                & 54.12              & 66.94             & 55.55                                        & 68.88                                        \\ \hline
84                                                                                  & 8,232                                                                                & 54.6               & 67.61             & 55.77                                        & 69.15                                        \\ \hline
\end{tabular}
\caption{Sample efficiency with changing relationships}
\label{tab:relation_efficiency}
\end{table}

In summary, these results demonstrate that a small number of training samples is
sufficient to obtain good performance, and that, when possible, the training
dataset should be diverse in terms of the kinds of relationships it represents.


%

\subsection{RQ5: FabricQA Performance in Context}
\label{subsubsec:fqa}

In the previous sections, we have shown the benefits of Relation
Coherence by measuring the improvement on extraction quality over
FabricQA. These results are valid as long as FabricQA is a
strong baseline. Otherwise, we would be improving over a suboptimal solution. In
this section, we demonstrate FabricQA is in fact a strong baseline.

\mypar{Datasets} Similar to previous sections, we use the SQuAD
1.1.~\cite{squad-1.0} benchmark to extract questions and ground truth answers.
However, instead of answering these questions over the passages provided in the
benchmark, we find the passage among the 35M passages from the indexed
Wikipedia. This approach is common among approaches that evaluate question
answering systems~\cite{drqa, multi_passage_bert}. In total, we answer 10,570
questions.

\mypar{Setup} Due to the spike in interest for building question answering
systems, many results are readily reported in the literature using the
dataset described above. Unlike in the original benchmark, there is not an
official leaderboard where results are posted. Instead, we rely on
results reported by previous work~\cite{asai2020learning} and on our own
deployment and evaluation to obtain the results. We need to rely on
previously reported results because the majority of approaches do not open
source their system or only open source a component, which is insufficient to
fully replicate the results. For this experiment, we configure FabricQA
to obtain the top-1 answer for each question and measure aggregated EM and F1.

\begin{table}[]
\center
\begin{tabular}{|l|l|l|}
\hline
\textbf{Models}                        & \textbf{EM} & \textbf{F1} \\ \hline
Multi-passage (Wang et al., 2019b)     & 53.0        & 60.9        \\ \hline
ORQA (Lee et al., 2019)                & 20.2        & -           \\ \hline
BM25+BERT (Lee et al., 2019)           & 33.2        & -           \\ \hline
Weaver (Raison et al., 2018)           & 42.3        & -           \\ \hline
RE3 (Hu et al., 2019)                  & 41.9        & 50.2        \\ \hline
MUPPET (Feldman \&El-Yaniv, 2019)      & 39.3        & 46.2        \\ \hline
BERTserini (Yang et al., 2019)         & 38.6        & 46.1        \\ \hline
DENSPI-hybrid (Seo et al., 2019)       & 36.2        & 44.4        \\ \hline
MINIMAL (Min et al., 2018)             & 34.7        & 42.5        \\ \hline
Multi-step Reasoner (Das et al., 2019) & 31.9        & 39.2        \\ \hline
Paragraph Ranker (Lee et al., 2018)    & 30.2        & -           \\ \hline
R3 (Wang et al., 2018a)                & 29.1        & 37.5        \\ \hline
DrQA (Chen et al., 2017)               & 29.8        & -           \\ \hline
PathRetriever (Asai et al., 2020)*     & 56.5        & 63.8        \\ \hline
SPARTA ( Zhao et al., 2021)*           & 59.3        & 66.5        \\ \hline
FabricQA                               & 54.3        & 61.7        \\ \hline
\end{tabular}
\caption{Question Answering results on Squad Open}
\label{tab:open_qa_model_quality}
\end{table}

\mypar{Results} Table \ref{tab:open_qa_model_quality} shows the results for
several approaches in the literature as well as for FabricQA.  There are only 2
approaches that achieve better quality than FabricQA. Of those,
PathRetriever\cite{asai2020learning} requires the corpus to contain hyperlinks
between documents and between paragraphs, which we do not require.
SPARTA\cite{zhao2020sparta} improves the passage retrieval using an expensive to
train vector-based sparse index, which is better at matching question and
passage than the token-based sparse index BM25, but at the cost of much lower
performance~\cite{zhu2021retrieving}. This is because SPARTA needs to encode
each token and passage using a learned function, and then it needs to precompute
the token and passage match score for each token in a vocabulary and passage in
the corpus. This requires a computationally expensive preprocessing
stage~\cite{zhu2021retrieving}. 

This experiment shows that FabricQA is a strong baseline and hence, it
demonstrates that the improvements of the Relation Coherence model are
significant.
 

\subsubsection{RQ6: Effect of Passage Size}
\label{subsubsec:rq5}

To explore the effect of passage size to the extraction quality,
we chunk documents into passages of different
sizes, N, that range from 10 to 200 tokens with strides of N/2. We report
results for 500 randomly sampled rows from each relation, resulting in a total
of 11903 rows.

\F\ref{fig:general_throughput_effect_passage} c) shows the results of this
experiment. Passages shorter than 60 tokens perform worse than those that are
larger than 60 ones, and the range between 80 to 140 tokens achieves the best
performance. Passages larger than 140 tokens harm performance. This loss in
performance is related to the limit size of RC models, that accept a maximum of
384 \emph{wordpiece} tokens~\cite{google_bert_code}---\ie roughly 200 tokens. If given
larger passages, the reader itself will chunk the passage and choose one based
on its own internal score. 

We note that the performance impact of the passage size can rest any benefits of
sophisticated ML models downstream the pipeline. This microbenchmark brings
attention to the importance of carefully designing these classes of systems
end-to-end.

\section{Related Work}
\label{sec:relatedwork}

To the best of our knowledge, \textsf{FabricQA-Extractor} is the first end-to-end
system proposed to solve information extraction tasks using question
answering over large collections of documents while exploiting relational
knowledge.  We covered related work with Table~\ref{table:landscape} in
Section~\ref{sec:background}. We complete the discussion here.

\mypar{Relation extraction using question answering} West et. al. \cite{kbc-qa}
use question answering to complete knowledge bases. 
The approach learns to ask questions for a missing
attribute of an entity and then aggregate the answers from those different
questions to get the possible values. Zhao et. al. \cite{zhao2020asking} also
feed diverse questions to an RC model to boost performance in relation
extraction. Das et. al \cite{das2018building} extends an RC model to construct a
knowledge graph from text to track and predict entity states.
All these approaches assume relationships and entity types are pre-defined and
cannot easily answer questions about new relationships without retraining.
('Requires Training per relationship' column in Table~\ref{table:landscape}),
Our work has no such constraints. 

Levy et. al. \cite{zero-shot-rel} uses reading comprehension to fill a missing
entity of a relation tuple. They extend the BIDAF \cite {bidaf} RC model to
report no answer if the passage does not imply the relation in the question.
This approach performs badly when a lot of passages are fed to a question
because exiting RC models are still poor in ruling out distractor passages and
ranking the related ones. Li et. al. \cite{li2019entity} reduces relation
extraction to a multi-turn QA task, likewise, they rely on the RC model to
report no answer for a passage. 
Qiu et. al \cite{qiu2018qa4ie} also use question answering to populate knowledge
bases, but they assume the right document is given instead of retrieving first.
All these approaches assume an input passage (or document)
and not a collection of documents ('Designed for large scale in
Table~\ref{table:landscape}'). And all propose a model instead of an end to end
system. 

\mypar{Open question answering} Open question answering (OpenQA) has become an
active topic since DrQA \cite{drqa} was proposed. It first introduced the
retriever-reader architecture and integrated a neural reading comprehension
model in OpenQA. Since then a lot of work has been done. \cite{wang2018r}
jointly trains retriever and reader using reinforcement learning.
\cite{yang2019end} proposed an end-to-end approach that combines the retriever
score and read score by hyperparameters. \cite{clark2017simple} proposed global
normalization to make the reader score is more comparable among passages. To
better capture the semantics between query term and passages, dense index
\cite{lee-etal-2019-latent} \cite{zhao2020sparta} is used to replace sparse ones
like BM25\cite{yang2019end} and TF-IDF\cite{drqa}. To answer more complex,
multi-hop questions, \cite{asai2020learning} exploits document hyperlinks
available to build a graph to collect more evidence passages.
\cite{xiong2020answering} does not require the hyperlinks and combines the
dense index and a recursive framework to solve the multi-hop questions. 
None of these systems performs information extraction tasks and hence none
exploits knowledge about the relational model.



\section{Conclusions}
\label{sec:conclusions}

We demonstrated the importance of exploiting relational knowledge to boost the
extraction quality from large collections of documents. We proposed
Relation Coherence, a novel model that exploits that information, and
we incorporated it as part of an end-to-end system, FabricQA-Extractor.
We demonstrated Relation Coherence's benefits on two large scale
datasets with tens of millions of passages using tens of thousands of questions.
We demonstrated the importance of building this class of information extraction
systems end to end, where components that are often left unattended---such as the
chunker---have a large impact on the overall quality. All in all, our work hints
at additional paths for the data management community to contribute to this
growing area.


\bibliographystyle{IEEEtran}
\balance
\bibliography{main}

\end{document}